\documentclass[12pt]{article}
\usepackage{amsfonts}
\usepackage{latexsym,amsmath}
\usepackage{amssymb,array}

\makeatletter \oddsidemargin 0in \evensidemargin 0in \textwidth 16cm
 \RequirePackage[dvips]{graphicx} \textheight 18cm
\newcommand{\al}{\alpha}
\newcommand{\om}{\omega}
\newcommand{\op}{\oplus}
\newcommand{\ga}{\gamma}
\newtheorem{t1}{Theorem}[section]

\newtheorem{c1}{Corollary}[section]

\newtheorem{ce1}{Counterexample}[section]
\newtheorem{d1}{Definition}[section]
\newtheorem{e1}{Example}[section]
\date{July, 2015}
\begin{document}
\title{On Stochastic Comparisons for Load-Sharing Series and Parallel Systems}
\author{Maxim Finkelstein\footnote{Corresponding author, e-mail:
FinkelM@ufs.ac.za}~ and Nil Kamal Hazra \\
\textit{\small{Department of Mathematical Statistics, University of the Free State,}}\\
\textit{\small{339 Bloemfontein 9300, South Africa }} }
\date{November, 2015}
\maketitle
\begin{abstract}
We study the allocation strategies for redundant components in the  
load-sharing series/parallel systems. We show that under the specified assumptions, the allocation of a redundant component to the stochastically weakest (strongest) component of a series (parallel) system is the best strategy to achieve its maximal reliability. The results have been studied under cumulative exposure model and for a general scenario as well. They have a clear intuitive meaning, however, the corresponding additional assumptions are not obvious, which can be seem from the proofs of our theorems.
\end{abstract}

{\bf \par Keywords \& Phrases:} Accelerated lifetime model; cumulative exposure model; hazard rate function; reversed hazard rate function; stochastic orders; virtual age

\section{Introduction}
One of the standard methods to enhance reliability of a system is to use redundancy in its structure (Barlow and Proschan~\cite{bp2}). The problem of optimal allocations of redundant components was addressed in numerous publications (cf. Boland \emph{et al.}~\cite{bep}, Brito \emph{et al.}~\cite{br1}, Misra and Misra~\cite{mm7}, Misra \emph{et al.}~(\cite{mmd2}, \cite{mmd22}), Romera \emph{et al.}~\cite{rvz2}, Vald\'{e}z and Zequeira~(\cite{vz71}, \cite{vz72}), Hazra and Nanda~(\cite{hn1}, \cite{hn}), Cha \emph{et al.}~\cite{cha}, Li \emph{et al.}~\cite{xli}, Yun and Cha~(\cite{yc}, \cite{yuncha}) to name a few). The specific case of redundancy, i.e., load sharing, had attracted much less attention (see, for example, Kapur and Lamberson~\cite{kl}, Keccecioglu~\cite{k}, Scheuer~\cite{se}, Shechner~\cite{s}, Lin \emph{et al.}~\cite{lcw}, Yinghui and Jing~\cite{yj}, Wang \emph{et al.}~\cite{whth}, Shao and Lamberson~\cite{sl2}, Liu~\cite{lh} and the references therein).
\\\hspace*{0.3 in} The load-sharing systems can be often encountered in practice, as a tool for decreasing electrical or mechanical stresses and therefore, increasing the corresponding reliability characteristics. The popular examples are: generators used in a power plant, CPU in a multiprocessor computer system, cables in a suspension bridge, valves or pumps used in a hydraulic system, bolts used to hold a mechanical system, etc.  However, most of the relevant papers in the literature deal only with the specific case when the components of the load sharing systems are described by the exponentially distributed lifetimes. There are very few references where the systems with components having arbitrary lifetime distributions have been considered (see for example, Liu~\cite{lh}, and Yun and Cha~\cite{yc}). The reason for that is that, for a general case, one must be able to recalculate the time that an item had spent in one regime (decreased load) after switching to another (full load).
\\\hspace*{0.3 in}In this paper we consider load-sharing series and parallel systems (formed by components having arbitrary lifetime distributions) when one of the components of the system shares  load with another component. We want to obtain the best allocation strategy that maximizes reliability of our system. Similar and more general settings for systems without load sharing are widely studied in the literature (see, e.g., Boland \emph{et al.}~\cite{bep}, Brito \emph{et al.}~\cite{br1}, Misra and Misra~\cite{mm7}, Misra \emph{et al.}~(\cite{mmd2}, \cite{mmd22}), Romera \emph{et al.}~\cite{rvz2}, Vald\'{e}z and Zequeira~(\cite{vz71}, \cite{vz72}))  However, to the best of our knowledge, there are no relevant results in the literature for the load sharing systems with arbitrary distributions of component's lifetimes.  The obtained in this paper results have a clear intuitive meaning similar to the case of 'ordinary' redundancy (without load sharing). On the other hand, due to the load sharing and subsequent recalculation of age a number of additional issues come into play that should be properly formulated and adequately described mathematically. This will be done in our paper.
\\\hspace*{0.3 in}For any continuous random variable $X$, denote by $F_X(\cdot)$ the cumulative distribution function, the probability density function by $f_X(\cdot)$ (whenever exists), the survival function by $\bar F_X(\cdot)$, the hazard (failure) rate function by $r_X(\cdot)$, and the reversed hazard rate function by $\tilde r_X(\cdot)$.
\\\hspace*{0.3 in}Consider a parallel system formed by two components, namely, $A$ and $B$. Assume that both $A$ and $B$ are initially sharing load, and after the failure of one component, the other one switches over to the full load condition. Without loss of generality, let the total load for  a system be $1$ and components $A$ and $B$ share $\al$ and $(1-\al)$ of it, respectively. Denote by $X$ and $Y$ the random variables representing the lifetimes of $A$ and $B$ under a full load and by $X^*$ and $Y^*$  those for the load sharing, respectively.  As the lifetime of a component in a partial load condition should be larger than that in the full load condition, we can use stochastic reasoning employed in accelerated life modeling (ALM) (see. e.g., Nelson~\cite{ne} and Finkelstein~\cite{f}). Thus, in accordance with the linear version of the ALM, we can write for both components  the following relationships
$$F_{X^*}(t)=F_X(g(\al)t)\quad \text{for all }t\geq 0$$
and
$$F_{Y^*}(t)=F_Y(h(1-\al)t)\quad \text{for all }t\geq 0,$$
where both $g(\cdot)$ and $h(\cdot)$ satisfy: ($i$) $0\leq g(\al)\leq 1 $ and $0\leq h(1-\al)\leq 1$, for all $\al\in[0,1]$, and ($ii$) both $g(\cdot)$ and $h(\cdot)$ are strictly increasing functions. The simplest specific case of these functions is when $g(\alpha)=\alpha$ and $h(1-\alpha)=1-\alpha$, however our assumptions allow for a more  practically important setting when dependence on the load sharing factor $\alpha$ is more general.
\\\hspace*{0.3 in}Suppose now that the component $A$ was operating under a partial load in $[0,u)$, and was switched to the full load at time $t=u$ due to the failure of the component $B$. Thus, we must recalculate the component's age $t$ before the switching and obtain its initial age after the switching to be called the \emph{virtual age} (see, e.g., Kijima~\cite{kijima} and Finkelstein~\cite{fink}). Denote it by $\om(u)$. The virtual age satisfies the following natural conditions: ($i$) $0\leq\om(u)\leq u$ for all $u\geq 0$, and ($ii$) $\om(\cdot)$ is an increasing function. Let us call $\{\om(\cdot), g(\cdot)\}$ the \emph{set of model functions} for $A$. For the `reverse' scenario, when the component $A$ fails before the component $B$, denote the virtual age of the component $B$ by $\ga(u)$ and the corresponding set of model functions  by $\{\ga(\cdot),h(\cdot)\}$.
Denote also the lifetime of the described load sharing system by $X\oplus Y$. Then the corresponding survival function is given by (cf. Yun and Cha~\cite{yc})
\begin{eqnarray}\label{eq12}
\bar F_{X\oplus Y}(t)=&&\bar F_X(g(\al)t)\bar F_Y(h(1-\alpha)t)\nonumber
\\&&+\int\limits_0^t \frac{\bar F_X(t-u+\om(u))}{\bar F_X(\om(u))}\bar F_X(g(\al)u)h(1-\al)f_Y(h(1-\al)u)du\nonumber
\\&&+\int\limits_0^t \frac{\bar F_Y(t-u+\ga(u))}{\bar F_Y(\ga(u))}\bar F_Y(h(1-\al)u)g(\al)f_X(g(\al)u)du.
  \end{eqnarray}
Indeed, the first term in the r.h.s corresponds to the case when both components did not fail in $[0,t)$, whereas the second and the third terms correspond to the cases when one of the components had failed and the other was functioning under the full load after that.
\\\hspace*{0.3 in} Consider now a series (resp. parallel) system formed by $n$ independent components with lifetimes $X_1,X_2,$ $\dots,X_n$. Let a redundant component for load sharing with the lifetime $Y$ be available for allocation to one of the components of the system. Then, the natural question is: how to allocate $Y$ in the system in order to maximize its reliability in a suitable stochastic sense?  Define, for $i=1,2,\dots,n$,
\begin{eqnarray*}
U_i=\min\{X_1,\dots, X_{i-1},X_i\oplus Y, X_{i+1},\dots, X_n\},
\end{eqnarray*}
and
\begin{eqnarray*}
V_i=\max\{X_1,\dots, X_{i-1},X_i\oplus Y, X_{i+1},\dots, X_n\},
\end{eqnarray*}
where the operation $X_i\oplus Y$ means that $Y$
is sharing load with $X_i$. Thus, $U_i$ (resp. $V_i$) represents the lifetime of a series (resp. parallel) system where the component $Y$ is sharing load with the $i$th component $X_i$.
Note also that in case of ordinary (not load sharing redundancy), this problem obviously, does not exist for parallel systems as all $n+1$  components will be functioning independently, whereas in the case of load sharing, the corresponding dependence takes place.
\\\hspace*{0.3 in} In order to achieve the optimal (maximal) reliability of our system, certain measures for comparison of reliability characteristics should be employed. It is well-known that stochastic ordering is a very useful tool for  comparing lifetimes. Many different types of stochastic orders have been developed and studied in the literature. For example, usual stochastic order compares two reliability functions, hazard rate order compares two failure rate functions, whereas the reversed hazard rate order compares two reversed hazard rate functions (see Shaked and Shanthikumar~\cite{shak1} for encyclopaedic information on stochastic orders). For the sake of completeness, we give the following definitions of stochastic orders that will be used in our paper.
\begin{d1}
Let $X$ and $Y$ be two continuous nonnegative random variables with respective supports $(l_X,u_X)$ and $(l_Y,u_Y)$, where $u_X$ and $u_Y$ may be positive infinite, and $l_X$ and $l_Y$ may be zero. Then, $X$ is said to be smaller than $Y$ in
\begin{enumerate}
\item hazard rate (hr) order, denoted as $X\leq_{hr}Y$, if
$$\frac{\bar F_Y(x)}{\bar F_X(x)}~~\text{is increasing in }x\in(0,\max(u_X,u_Y)),$$
which can equivalently be written as
$$r_X(x)\geq r_Y(x),~~\text{where defined};$$
\item reversed hazard rate (rhr) order, denoted as $X\leq_{rhr}Y$, if
$$\frac{ F_Y(x)}{ F_X(x)}~~\text{is increasing in }x\in(\min(l_X,l_Y), \infty),$$
which can equivalently be written as
$$\tilde r_X(x)\leq \tilde r_Y(x),~~\text{where defined};$$
\item usual stochastic (st) order, denoted as $X\leq_{st}Y$, if
$$\bar F_X(x)\leq \bar F_Y(x)~~\text{for all}~t\in(0,\infty).$$
\end{enumerate}
\end{d1}
\hspace*{0.3 in}The following diagram shows the chain of implications among the stochastic orders as discussed above.
\\\hspace*{1.7 in}$X\leq_{hr}Y$
\\\hspace*{1.7 in}$ ~~~~~~~~~~~~\searrow$
\\\hspace{1.7 in} $~~~~~~~~~~~~~~~~~~~~~~~~~~~~~~~~~~~~~~~~~~~~~~~~~~X\leq_{st}Y$.

\hspace{1.7 in}~~~~~~~~$\nearrow$

\hspace{1.5 in}$X\leq_{rhr}Y$
\vspace*{0.17 in}

Thus the hazard rate order and the reversed hazard rate order are stronger than usual stochastic order. Throughout the paper, increasing and decreasing, as usually means non-increasing and non-decreasing, respectively. The random variables considered in this paper are all nonnegative. For convenience of notation, we write $Z=\min\{X_3,X_4,\dots,X_n\}$ and $W=\max\{X_3,X_4,$ $\dots,X_n\}$.
\\\hspace*{0.3 in}The rest of the paper is organized as follows. In Section~\ref{se1}, we consider allocation strategies for load-sharing series (resp. parallel) systems. We compare different variants of these systems with respect to the usual stochastic order  and under the assumption of the cumulative exposure model. We generalize these results in Section~\ref{se2}. 
Finally, the concluding remarks are given in Section~\ref{se3}.
\section{Stochastic Comparisons under the Cumulative Exposure Model}\label{se1}
As was stated in the Introduction, when a component of a load sharing system is switched from the partial load to a full load, its age should be recalculated. The  cumulative exposure model or its equivalents (see, e.g., Nelson~\cite{ne} and Finkelstein~\cite{f}) that is widely used in accelerated life testing, is a popular and efficient way to do so.  In this section, we also use this model for the corresponding age recalculation. Consider the load-sharing system as discussed in (\ref{eq12}). In accordance with the reasoning similar to the cumulative exposure model, e.g., for the component $X$, we have the following relation for the virtual age after switching: 
\[
 F_{X^*}(t)=F_{X}(g(\al)t)=F_{X}(\om(t)), 
 \]
which immediately gives us $\om(t)=g(\al)t$ for all $t\geq 0$. Thus, the virtual age of a component after the switching to the full load is smaller than the age before switching and, in accordance with our assumptions, is given by the linear function. Note that this function also defines the scale transformation in the argument of the corresponding distribution function under partial load (ALM). In a similar way we can also calculate the virtual age $\ga(t)$ for the component $Y$. Note that it was also implicitly assumed that the 'form' of the corresponding remaining lifetime distribution function for the full load does not depend on when the switching to the full load had occurred (only initial/virtual age differs i.e., the baseline distribution is the same).

In what follows in this section, we will formulate and analyze several practically important stochastic comparisons of interest for the case of the linear virtual age described above. Note that, this linear virtual age is, in fact, a consequence of our assumption that the accumulated exposure model holds. In the next section, we will not rely on this assumption and, similar to (\ref{eq12}) will consider the case of general, not necessarily linear virtual ages. For the sake of presentation, we will omit now all the proofs that are just specific cases of general results of the next section, for which the detailed proofs will be given. Thus, the contents of the current section can have practical importance, whereas the results of Section 3 are more theoretical. 
\\\hspace*{0.3 in}Suppose that we have two different components, and one  redundant component that can be used in a load sharing scenario with any of the two components.  Then, the following theorem holds (see the proof of  Theorem~\ref{th31}).
\begin{t1}\label{th11}
Let $\{\om_1(\cdot),g_1(\cdot)\}$, $\{\om_2(u),g_2(\cdot)\}$ and $\{\ga(\cdot),h(\cdot)\}$ be the sets of model functions for $X_1$, $X_2$ and $Y$, respectively. Suppose that the following conditions hold.
\begin{itemize}
\item [$(i)$]$X_1\geq_{st}X_2$.
\item [$(ii)$] $\om_1(u)=g_1(\al)u\leq g_2(\al)u=\om_2(u)$ and $\ga(u)=h(1-\al)u$, for all $0\leq \al\leq 1$ and $u\geq 0$.
\end{itemize}
Then, $X_1\op Y\geq_{st}X_2\op Y$.$\hfill\Box$
\end{t1}

For convenience, let us say that one component (system) is stronger (weaker) than another if its lifetime is larger (smaller) in the sense of the usual stochastic ordering. If another ordering is used, then we will add the corresponding description where necessary. Thus this theorem states that if the redundant component for load sharing is allocated to the stronger component, then the system with load sharing will be also stronger.  This result can be, of course, intuitively anticipated as the same holds for ordinary, not load sharing systems with active (hot) or standby (cold) redundant component. However, an important feature of our result is that additionally we need an ordering between virtual ages. Otherwise, the proposed ordering does not necessarily hold as the following counterexample shows.
 \begin{ce1}\label{cde}
 Let $X_1$ and $X_2$ be two independent random variables representing the lifetimes of two components with failure rates $1$ and $1.2$, respectively. Further, let $Y$ be another random variable representing the lifetime of a redundant component with the failure rate $2$. Assume that $X_1$, $X_2$ and $Y$ are independent. Let $\om_1(u)=g_1(\al)u=0.5 u$, $\om_2(u)=g_2(\al)u=0.25 u$ and $\ga(u)=h(1-\al)u=0.5u$, for all $u\geq 0$. Then, $X_1\geq_{st}X_2$ but $\om_1(u)\nleq \om_2(u)$. Denote: $e_0(t)=\bar F_{X_1\op Y}(t)-\bar F_{X_2\op Y}(t)$. Then, for all $t\geq 0$,
 \begin{eqnarray*}
 e_0(t)&=&\int\limits_{0}^t\left[\bar F_{X_1}(t-u+\om_1(u))-\bar F_{X_2}(t-u+\om_2(u))\right]dF_Y(h(1-\al)u)
 \\&&+\int\limits_{0}^t\left[\bar F_{X_1}(g_1(\al)u)-\bar F_{X_2}(g_2(\al)u)\right]d\bar F_Y(t-u+\ga(u))
 \\&=&\int\limits_{0}^te^{-u}\left[e^{-(t-u+0.5u)}-e^{-1.2(t-u+0.25u)}\right]du
 +\int\limits_{0}^t e^{-2(t-u+0.5u)}\left[e^{-0.5u}-e^{-0.3u}\right]du.
 \end{eqnarray*}
 \end{ce1}
It is easy to prove that $e_0(t)$ is not always nonnegative, and hence $X_1\op Y\ngeq_{st}X_2\op Y$.$\hfill\Box$
\\\hspace*{0.3 in}The following example illustrates the result given in the above theorem.
\begin{e1}
Let $X_1$ and $X_2$ be two independent random variables representing the lifetimes of two components with failure rates $\lambda_1$ and $\lambda_2$, respectively, where $0<\lambda_1\leq \lambda_2$. Furthermore, let $Y$ be another random variable representing the lifetime of a redundant component with the failure rate $\mu~(>0)$. Assume that $X_1$, $X_2$ and $Y$ are independent. Let $\om_1(u)=g_1(\al)u=\al^2 u$, $\om_2(u)=g_2(\al)u=\al u$ and $\ga(u)=h(1-\al)u=(1-\al)u$, for all
$0\leq \al\leq 1$ and $u\geq 0$. Then all conditions given in Theorem~\ref{th11} are satisfied. Hence, $X_1\op Y\geq_{st}X_2\op Y$.
\end{e1}
\hspace*{0.3 in}In the following theorem we compare two series systems where each system has a component which shares partial load with a redundant component. We show that if we allocate the redundancy to the weakest component of a series system then the resulting system becomes optimal in the sense of usual stochastic order (see the proof of Theorem~\ref{th32}).
\begin{t1}\label{th22}
Let $\{\om_1(\cdot),g_1(\cdot)\}$, $\{\om_2(u),g_2(\cdot)\}$ and $\{\ga(\cdot),h(\cdot)\}$ be the sets of model functions for $X_1$, $X_2$ and $Y$, respectively. Suppose that the following conditions hold.
\begin{itemize}
\item [$(i)$] $X_1\leq_{hr}X_2$.
\item [$(ii)$] $\om_1(u)=g_1(\al)u\leq g_2(\al)u=\om_2(u)$ and $\ga(u)=h(1-\al)u$, for all $0\leq \al\leq 1$ and $u\geq 0$.

\end{itemize}
Then, $U_1\geq_{st}U_2$.$\hfill\Box$
\end{t1}
\hspace*{0.3 in}The following corollary immediately follows from Theorem~\ref{th22}.
\begin{c1}
Let $\{\om_i(\cdot), g_i(\cdot)\}$ be the set of model functions for $X_i$, $i=1,2,\dots,n$, and $\{\ga(\cdot), h()\cdot\}$ be that for $Y$.
Suppose that the following conditions hold.
\begin{itemize}
\item [$(i)$] $X_1\leq_{hr}X_2\leq_{hr}\dots\leq_{hr}X_n$.
\item [$(ii)$] $\om_1(u)=g_1(\al)u\leq \om_2(u)=g_2(\al)u\leq\dots \leq \om_n(u)=g_n(\al)u$ and $\ga(u)=h(1-\al)u$, for all $0\leq \al\leq 1$ and $u\geq 0$.

\end{itemize}
Then, $U_1\geq_{st}U_2\geq_{st}\dots\geq_{st}U_n$.$\hfill\Box$
\end{c1}
\hspace*{0.3 in} Theorem~\ref{th22} can be  illustrated by the following example.
\begin{e1}
Let $X_1,X_2,\dots,X_n$ be $n$ independent random variables representing the lifetimes of $n$ components with failure rates $\lambda_1,\lambda_2,\dots,\lambda_n$, respectively, where all $\lambda_i$'s are positive with $\lambda_1\geq \lambda_2$. Further, let $Y$ be another random variable representing the lifetime of a redundant component with the failure rate $\mu~(>0)$. Assume that all $X_i$'s and $Y$ are independent. Let $\om_1(u)=g_1(\al)u=\al^2 u$, $\om_2(u)=g_2(\al)u=\al u$ and $\ga(u)=h(1-\al)u=(1-\al)u$, for all
$0\leq \al\leq 1$ and $u\geq 0$. It can be seen that all the conditions given in Theorem~\ref{th22} are satisfied. Hence, $U_1\geq_{st}U_2$.
\end{e1}
\hspace*{0.3 in} The intuitive meaning of Theorem~\ref{th22} is also quite clear: to allocate the redundant component to the weakest component of the system. However, distinct from the previous theorem and additional to 'intuitive reasoning' the weakest component is understood in the sense of the hazard rate ordering. This is an important observation that could not be foreseen without a proper proof. The following counterexample shows that the condition $X_1\leq_{hr}X_2$ given in Theorem~\ref{th22} can not be replaced by $X_1\leq_{st}X_2$.
\begin{ce1}\label{ceq1}
Fix $n=2$. Let $X_1$, $X_2$ and $Y$ be independent random variables with cumulative distribution functions given by
 \begin{align*}
  F_{X_1}(t)=
  \left \{
      \begin{array}{ll}
            \frac{t}{3}, &\text{for}~\; 0\leq t \leq 3 \\
            1, & \text{for}~\;t \geq 3,\\
       \end{array}
 \right.
\end{align*}
 \begin{align*}
  F_{X_2}(t)=
  \left \{
      \begin{array}{lll}
            \frac{t^2}{3}, &\text{for}~\; 0\leq t \leq 1 \\
            \frac{t^2+3}{12}, &\text{for}~\; 1\leq t \leq 3\\
            1, & \text{for}~\;t \geq 3\\
       \end{array}
 \right.
\end{align*}
and
$$F_Y(t)=1-e^{-3t}, \quad \text{for}~ t>0,$$
respectively.
Then, it is easy to verify that $X_1\leq_{st}X_2$ but $X_1\nleq_{hr}X_2$. Further, let $\om_1(u)=g_1(\al)u=0.01 u$, $\om_2(u)=g_2(\al)u=0.1 u$ and $\ga(u)=h(1-\al)u=0.9 u$. Note that condition ($ii$) given in Theorem~\ref{th22} is satisfied. Denote: $e_1(t)=\bar F_{U_1}(t)-\bar F_{U_2}(t)$. Then, for all $0\leq t\leq 1$,
\begin{eqnarray*}
e_1(t)&=&\bar F_Y(t)\left[\bar F_{X_2}(t)-\bar F_{X_1}(t)\right]
\\&&+\int\limits_{0}^t \left[\bar F_{X_2}(t)\bar F_{X_1}(g_1(\al)u)-\bar F_{X_1}(t)\bar F_{X_2}(g_2(\al)u)\right]d\bar F_Y(t-u+\ga(u))
\\&&+\int\limits_{0}^t \left[\bar F_{X_2}(t)\bar F_{X_1}(t-u+\om_1(u))-\bar F_{X_1}(t)\bar F_{X_2}(t-u+\om_2(u))\right]d F_Y(h(1-\al)u)
\\&=&e^{-3t}(t/3-t^2/3)
\\&&+\int\limits_0^t 0.3\left[\left(1-\frac{t^2}{3}\right)\left(1-\frac{0.01u}{3}\right)-\left(1-\frac{t}{3}\right)\left(1-\frac{0.01u^2}{3}\right)\right]e^{-3(t-u+0.9u)}du
\\&&+\int\limits_0^t 2.7\left[\left(1-\frac{t^2}{3}\right)\left(1-\frac{t-u+0.01u}{3}\right)-\left(1-\frac{t}{3}\right)\left(1-\frac{(t-u+0.1u)^2}{3}\right)\right]e^{-2.7u}du.
\end{eqnarray*}
It can be verified that $e_1(t)$ is not always nonnegative. Thus, $U_1\ngeq_{st}U_2$.$\hfill\Box$
\end{ce1}
\hspace*{0.3 in} We are turning now to analysis of parallel load sharing systems. In the following theorem, we compare two load-sharing parallel systems with $n$ initial components. We show that the best strategy to get an optimal parallel system (in the sense of usual stochastic order) is to allocate the redundant component for the load sharing to the strongest component of the system. (See the proof of Theorem~\ref{th33}). Note that this can be considered as a new problem as in the case of ordinary (not load sharing) redundancy, obviously, it does not matter to which component to allocate it.
\begin{t1}\label{th23}
Let $\{\om_1(\cdot),g_1(\cdot)\}$, $\{\om_2(u),g_2(\cdot)\}$ and $\{\ga(\cdot),h(\cdot)\}$ be the sets of model functions for $X_1$, $X_2$ and $Y$, respectively. Suppose that the following conditions hold.
\begin{itemize}
\item [$(i)$] $X_1\geq_{rhr}X_2$.
\item [$(ii)$] $\om_1(u)=g_1(\al)u\leq g_2(\al)u=\om_2(u)$ and $\ga(u)=h(1-\al)u$, for all $0\leq \al\leq 1$ and $u\geq 0$.

\end{itemize}
Then, $V_1\geq_{st}V_2$.$\hfill\Box$
\end{t1}
\hspace*{0.3 in}Thus, to obtain a more reliable system, it is not sufficient that the components should be ordered in the sense of the usual stochastic order. It should be a stronger reversed hazard rate order, which is an interesting observation.

The following corollary directly follows from Theorem~\ref{th23}.
\begin{c1}
Let $\{\om_i(\cdot), g_i(\cdot)\}$ be the set of model functions for $X_i$, $i=1,2,\dots,n$, and $\{\ga(\cdot), h()\cdot\}$ be that for $Y$.
Suppose that the following conditions hold.
\begin{itemize}
\item [$(i)$] $X_1\geq_{rhr}X_2\geq_{rhr}\dots\geq_{rhr}X_n$.
\item [$(ii)$] $\om_1(u)=g_1(\al)u\leq \om_2(u)=g_2(\al)u\leq\dots \leq \om_n(u)=g_n(\al)u$ and $\ga(u)=h(1-\al)u$, for all $0\leq \al\leq 1$ and $u\geq 0$.

\end{itemize}
Then, $V_1\geq_{st}V_2\geq_{st}\dots\geq_{st}V_n$.$\hfill\Box$
\end{c1}
\hspace*{0.3 in}  The following example illustrates the result of Theorem~\ref{th23}.
\begin{e1}
Let $X_1, X_2,\dots,X_n$ be $n$ independent random variables representing the lifetimes of $n$ components. Further, let $X_i$ have the survival function given by $\bar F_{X_i}(t)=(1+(tk_i-\sigma_i)/\sigma_i)^{-1/{k_i}}$, $t>\sigma_i/k_i>0$, $k_i>0$, for $i=1,2,\dots,n$. Suppose that $k_1\geq k_2$.  Assume that all $X_i$'s and $Y$ are independent. Let $\om_1(u)=g_1(\al)u=\al^2 u$, $\om_2(u)=g_2(\al)u=\al u$ and $\ga(u)=h(1-\al)u=(1-\al)u$, for all
$0\leq \al\leq 1$ and $u\geq 0$. Note that all conditions of Theorem~\ref{th23} are satisfied. Hence, $V_1\geq_{st}V_2$.$\hfill\Box$ 
\end{e1}
\hspace*{0.3 in} The following important counterexample shows that the condition $X_1\geq_{rhr}X_2$ given in Theorem~\ref{th23} can not be replaced by $X_1\geq_{st}X_2$.
\begin{ce1}\label{ceq2}
Fix $n=2$. Let $X_1$, $X_2$ and $Y$ be independent random variables with cumulative distribution functions given by
 \begin{align*}
  F_{X_1}(t)=
  \left \{
      \begin{array}{lll}
            \frac{t^2}{3}, &\text{for}~\; 0\leq t \leq 1 \\
            \frac{t^2+3}{12}, &\text{for}~\; 1\leq t \leq 3\\
            1, & \text{for}~\;t \geq 3,\\
       \end{array}
 \right.
\end{align*}
 \begin{align*}
  F_{X_2}(t)=
  \left \{
      \begin{array}{ll}
            \frac{t}{3}, &\text{for}~\; 0\leq t \leq 3 \\
            1, & \text{for}~\;t \geq 3\\
       \end{array}
 \right.
\end{align*}
and
$$F_Y(t)=1-e^{-3t}, \quad \text{for}~ t>0,$$
respectively.
Then, it is easy to verify that $X_1\geq_{st}X_2$ but $X_1\ngeq_{rhr}X_2$. Further, let $\om_1(u)=g_1(\al)u=0.01 u$, $\om_2(u)=g_2(\al)u=0.1 u$ and $\ga(u)=h(1-\al)u=0.9 u$. Note that condition ($ii$) given in Theorem~\ref{th23} is satisfied. Denote $e_2(t)=F_{V_2}(t)- F_{V_1}(t)$. Then, for all $1\leq t\leq 3$,
\begin{eqnarray*}
e_2(t)&=&\int\limits_{0}^t \left[ F_{X_1}(t)F_{X_2}(g_2(\al)u)- F_{X_2}(t) F_{X_1}(g_1(\al)u)\right]d\bar F_Y(t-u+\ga(u))
\\&&+\int\limits_{0}^t \left[ F_{X_1}(t) F_{X_2}(t-u+\om_2(u))- F_{X_2}(t) F_{X_1}(t-u+\om_1(u))\right]d F_Y(h(1-\al)u)
\\&=&\int\limits_0^t 0.3\left[\left(\frac{t^2+3}{12}\right)\left(\frac{0.01u}{3}\right)-\left(\frac{t}{3}\right)\left(\frac{0.0001u^2+3}{12}\right)\right]e^{-3(t-u+0.9u)}du
\\&&+\int\limits_0^t 2.7\left[\left(\frac{t^2+3}{12}\right)\left(\frac{t-u+0.1u}{3}\right)-\left(\frac{t}{3}\right)\left(\frac{(t-u+0.01u)^2+3}{12}\right)\right]e^{-2.7u}du.
\end{eqnarray*}
It is easy to verify that $e_2(t)$ is not always nonnegative. Thus, $V_1\ngeq_{st}V_2$. $\hfill\Box$
\end{ce1}
\section{General Scenario}\label{se2}
In the previous section the assumption of the cumulative exposure model was used to obtain the corresponding virtual ages after switching. Therefore, combined with the scale transformation assumption under the partial load (ALM), it had resulted in the linear virtual age. As was already mentioned, we will not assume in this section that the  cumulative exposure model holds and will consider general forms of the virtual age functions. However, as previously we will  assume that the 'form' of the corresponding remaining lifetime distribution function for the full load does not depend on when the switching to the full load had occurred (only initial/virtual age differs i.e., the baseline distribution is the same). Thus the theorems of this section are generalizations of the corresponding theorems of the previous section where more practical results where presented. Our presentation of the following results are more formal. it should be noted that our theorems here employ quite a number of additional assumptions, however their probabilistic meaning is quite clear and can be easily interpreted.
 
 The following theorem is a generalization of Theorem~\ref{th11} 
\begin{t1}\label{th31}
Let $\{\om_1(\cdot),g_1(\cdot)\}$, $\{\om_2(u),g_2(\cdot)\}$ and $\{\ga(\cdot),h(\cdot)\}$ be the sets of model functions for $X_1$, $X_2$ and $Y$, respectively. Suppose that the following conditions hold.
\begin{itemize}
\item [$(i)$] $g_1(\al)\leq g_2(\al)$ for all $0\leq \al\leq 1$, and $\om_1(u)\leq \om_2(u)$ for all $u\geq 0$.
\item [$(ii)$] $u-\ga (u)$ and $\ga(u)-h(1-\al)u$ are increasing in $u\geq 0$, for all $0\leq \al\leq 1$.
\item [$(iii)$]$X_1\geq_{hr}X_2$, and $X_1$ or $X_2$ has log-concave survival function.
\item [$(iv)$] $Y$ has log-concave survival function.
\end{itemize}
Then, $X_1\op Y\geq_{st}X_2\op Y$.
\end{t1}
{\bf Proof:} We only prove the result when $X_1$ has log-concave survival function. The result follows similarly for the other case. 
Note that
$$\bar F_{X_1\op Y}(t)-\bar F_{X_2\op Y}(t)=l_1(t)+l_2(t),$$
where
\begin{eqnarray*}
l_1(t)&=&\int\limits_0^t h(1-\al)f_Y(h(1-\al)u)\left[\frac{\bar F_{X_1}(t-u+\om_1(u))}{\bar F_{X_1}(\om_1(u))}\bar F_{X_1}(g_1(\al)u)\right.
\\&&\left.-\frac{\bar F_{X_2}(t-u+\om_2(u))}{\bar F_{X_2}(\om_2(u))}\bar F_{X_2}(g_2(\al)u)\right]du
\end{eqnarray*}
and
\begin{eqnarray*}
l_2(t)&=&\bar F_{Y}(h(1-\al)t)\left[\bar F_{X_1}(g_1(\al)t) -\bar F_{X_2}(g_2(\al)t)\right]
\\&&-\int\limits_0^t \bar F_Y(h(1-\al)u)\frac{\bar F_{Y}(t-u+\ga(u))}{\bar F_{Y}(\ga(u))}d\left[\bar F_{X_1}(g_1(\al)u)-\bar F_{X_2}(g_2(\al)u)\right]\nonumber
\\&=&\int\limits_0^t \left[\bar F_{X_1}(g_1(\al)u)-\bar F_{X_2}(g_2(\al)u)\right]d\left[ \bar F_Y(h(1-\al)u)\frac{\bar F_{Y}(t-u+\ga(u))}{\bar F_{Y}(\ga(u))}\right].
\end{eqnarray*}
To prove the result it suffices to show that both $l_1(t)$ and $l_2(t)$ are nonnegative. 
Note that
\begin{eqnarray*}
l_1(t)
&\geq & \int\limits_0^t h(1-\al)f_Y(h(1-\al)u)\bar F_{X_2}(g_2(\al)u)\left[\frac{\bar F_{X_1}(t-u+\om_1(u))}{\bar F_{X_1}(\om_1(u))}\right.
\\&&\left.-\frac{\bar F_{X_2}(t-u+\om_2(u))}{\bar F_{X_2}(\om_2(u))}\right]du
\\&\geq & \int\limits_0^t h(1-\al)f_Y(h(1-\al)u)\bar F_{X_2}(g_2(\al)u)\left[\frac{\bar F_{X_1}(t-u+\om_2(u))}{\bar F_{X_1}(\om_2(u))}\right.
\\&&\left.-\frac{\bar F_{X_2}(t-u+\om_2(u))}{\bar F_{X_2}(\om_2(u))}\right]du
\\&\geq &0,
\end{eqnarray*}
where the first inequality follows from the fact that $X_1\geq_{hr}X_2$ and $g_1(\al)\leq g_2(\al)$. The second inequality holds because $X_1$ has log-concave survival function, and $\om_1(u)\leq \om_2(u)$, whereas the third inequality follows from $X_1\geq_{hr}X_2$.
Further, for all $u\geq 0$, we have
\begin{eqnarray*}
\frac{d}{du}\left(\frac{\bar F_Y(h(1-\al)u)}{\bar F_{Y}(\ga(u))}\right)&=&\frac{\bar F_Y(h(1-\al)u)}{\bar F_{Y}(\ga(u))}\left[\ga'(u)\frac{f_Y(\ga(u))}{\bar F_{Y}(\ga(u))}-h(1-\al)\frac{f_Y(h(1-\al)u)}{\bar F_{Y}(h(1-\al)u)}\right]
\\&\geq &h(1-\al)\frac{\bar F_Y(h(1-\al)u)}{\bar F_{Y}(\ga(u))}\left[\frac{f_Y(\ga(u))}{\bar F_{Y}(\ga(u))}-\frac{f_Y(h(1-\al)u)}{\bar F_{Y}(h(1-\al)u)}\right]
\\&\geq &0,
\end{eqnarray*}
where the first inequality follows from the fact that both $\ga(u)$ and
$\ga(u)-h(1-\al)u$ are increasing in $u\geq 0$. The second inequality holds because $\ga(u)\geq h(1-\al)u$, for all $u\geq 0$, and $Y$ has log-concave survival function. Thus,
\begin{eqnarray}\label{eq5}
\frac{\bar F_Y(h(1-\al)u)}{\bar F_{Y}(\ga(u))}\quad \text{is increasing in}\;u\geq 0.
\end{eqnarray}
Since, $u-\ga (u)$ is increasing in $u\geq 0$, we have that
\begin{eqnarray}\label{eq6}
\bar F_{Y}(t-u+\ga(u))\quad\text{is increasing in}\;u\geq 0.
\end{eqnarray}
Thus, from (\ref{eq5}) and (\ref{eq6}), we get that
\begin{eqnarray}\label{eq7}
 \bar F_Y(h(1-\al)u)\frac{\bar F_{Y}(t-u+\ga(u))}{\bar F_{Y}(\ga(u))}\quad \text{is increasing in}\;u\geq 0.
\end{eqnarray}
Again, $X_1\geq_{hr} X_2$ and $g_1(\al)\leq g_2(\al)$ imply that, for all $u\in[0,t]$,
\begin{eqnarray}\label{eq11}
\bar F_{X_1}(g_1(\al)u)-\bar F_{X_2}(g_2(\al)u)\geq 0.
\end{eqnarray}
Thus, on using (\ref{eq7}) and (\ref{eq11}), we have that $l_2(t)\geq 0$, and hence the result is proved.$\hfill\Box$
\\\hspace*{0.3 in}In the following theorem we compare two load-sharing series systems. We show that allocation of the redundant component to the stochastically weakest component of the system is the best strategy (in the sense of usual stochastic order) to get the optimal series system (see Theorem 2.2).
\begin{t1}\label{th32}
Let $\{\om_1(\cdot),g_1(\cdot)\}$, $\{\om_2(u),g_2(\cdot)\}$ and $\{\ga(\cdot),h(\cdot)\}$ be the sets of model functions for $X_1$, $X_2$ and $Y$, respectively. Suppose that $(i)$ and $(ii)$, and any one among $(iii)$, $(iv)$, $(v)$, $(vi)$ hold.
\begin{itemize}
\item [$(i)$] $X_1\leq_{hr}X_2$ and $g_1(\al)\leq g_2(\al)$, for all $0\leq \al\leq 1$.
\item [$(ii)$] $Y$ has log-concave survival function, and $u-\ga (u)$ and $\ga(u)-h(1-\al)u$ are increasing in $u\geq 0$, for all $0\leq \al\leq 1$.
\item [$(iii)$] $X_1$ has log-concave survival function, and $ \max\{g_1(\al )u,\om_1(u)\}\leq \om_2(u)$, for all $u\geq 0$ and $0\leq \al\leq 1$.
\item [$(iv)$] $X_1$ has log-convex survival function, and $g_1(\al )u\leq \om_2(u)\leq \om_1(u)$, for all $u\geq 0$ and $0\leq \al\leq 1$.
\item [$(v)$] $X_2$ has log-concave survival function, and $g_1(\al )u\leq \om_1(u)\leq \om_2(u)$, for all $u\geq 0$ and $0\leq \al\leq 1$.
\item [$(vi)$] $X_2$ has log-convex survival function, and $ \max\{g_1(\al )u,\om_2(u)\}\leq \om_1(u)$, for all $u\geq 0$ and $0\leq \al\leq 1$.
\end{itemize}
Then, $U_1\geq_{st}U_2$.
\end{t1}
{\bf Proof:} We only prove the result under conditions $(i)$, $(ii)$ and $(iii)$. The result follows similarly for the other cases.
Note that
\begin{eqnarray*}
\bar F_{U_1}(t)=\bar F_{X_1\op Y}(t)\bar F_{X_2}(t)\bar F_Z(t)
\end{eqnarray*}
and
\begin{eqnarray*}
\bar F_{U_2}(t)=\bar F_{X_2\op Y}(t)\bar F_{X_1}(t)\bar F_Z(t).
\end{eqnarray*}
Writing $\Im(t)=\bar F_{U_1}(t)-\bar F_{U_2}(t)$, we have
\begin{eqnarray*}
\Im(t)=k_1(t)+k_2(t),
\end{eqnarray*}
where
\begin{eqnarray*}
k_1(t)&=&\bar F_Z(t)\bar F_{Y}(h(1-\al)t)\left[\bar F_{X_2}(t)\bar F_{X_1}(g_1(\al)t) -\bar F_{X_1}(t)\bar F_{X_2}(g_2(\al)t)\right]
\\&&+\int\limits_0^t \bar F_Z(t)\bar F_Y(h(1-\al)u)\frac{\bar F_{Y}(t-u+\ga(u))}{\bar F_{Y}(\ga(u))}\left[\bar  F_{X_2}(t)g_1(\al)f_{X_1}(g_1(\al)u)\right.
\\&&\left.-\bar F_{X_1}(t)g_2(\al)f_{X_2}(g_2(\al)u)\right]du
\\&=&\bar F_Z(t)\bar F_{Y}(t)\left[\bar F_{X_2}(t)-\bar F_{X_1}(t)\right]
\\&&+\int\limits_0^t \bar F_Z(t)\left[\bar  F_{X_2}(t)\bar F_{X_1}(g_1(\al)u)-\bar F_{X_1}(t)\bar F_{X_2}(g_2(\al)u)\right]d\left[\bar F_Y(h(1-\al)u)\frac{\bar F_{Y}(t-u+\ga(u))}{\bar F_{Y}(\ga(u))}\right]
\end{eqnarray*}
and
\begin{eqnarray*}
k_2(t)=&&\int\limits_0^t \bar F_Z(t)h(1-\al)f_Y(h(1-\al)u)\left[\frac{\bar F_{X_1}(t-u+\om_1(u))}{\bar F_{X_1}(\om_1(u))}\bar F_{X_1}(g_1(\al)u)\bar F_{X_2}(t)\right.
\\&&\left.-\frac{\bar F_{X_2}(t-u+\om_2(u))}{\bar F_{X_2}(\om_2(u))}\bar F_{X_2}(g_2(\al)u)\bar F_{X_1}(t)\right]du.
\end{eqnarray*}
To prove the result, it is sufficient to show that both $k_1(t)$ and $k_2(t)$  are nonnegative.
From ($i$), we have
\begin{eqnarray}\label{eq15}
\bar F_{X_2}(t)\bar F_{X_1}(g_1(\al)t) -\bar F_{X_1}(t)\bar F_{X_2}(g_2(\al)t)\geq 0
\end{eqnarray}
and
\begin{eqnarray}\label{eq17}
\bar F_{X_2}(t)-\bar F_{X_1}(t)\geq 0.
\end{eqnarray}
 Further, ($ii$) implies that (see the proof of Theorem~\ref{th31})
 \begin{eqnarray}\label{eq16}
 \frac{\bar F_Y(t-u+\ga(u))}{\bar F_Y(\ga(u))}\bar F_Y(h(1-\al)u\;\;\text{is increasing in }u\geq 0.
\end{eqnarray}
Thus, on using (\ref{eq15}), (\ref{eq17}) and (\ref{eq16}), we have that $k_1(t)\geq0$.
Again, ($iii$) implies that
\begin{eqnarray}\label{eq3}
\frac{\bar F_{X_1}(t-u+\om_1(u))}{\bar F_{X_1}(\om_1(u))}
\geq\frac{\bar F_{X_1}(t-u+\om_2(u))}{\bar F_{X_1}(\om_2(u))},
\end{eqnarray}
which can equivalently be written as
\begin{eqnarray*}
\frac{\bar F_{X_1}(t-u+\om_1(u))}{\bar F_{X_1}(\om_1(u))}\bar F_{X_1}(g_1(\al)u)\bar F_{X_2}(t)
&\geq &\frac{\bar F_{X_1}(t-u+\om_2(u))}{\bar F_{X_1}(\om_2(u))}\bar F_{X_1}(g_1(\al)u)\bar F_{X_2}(t)
\\&\geq & \frac{\bar F_{X_2}(t-u+\om_2(u))}{\bar F_{X_1}(\om_2(u))}\bar F_{X_1}(g_1(\al)u)\bar F_{X_1}(t)
\\&\geq &\frac{\bar F_{X_2}(t-u+\om_2(u))}{\bar F_{X_2}(\om_2(u))}\bar F_{X_2}(g_2(\al)u)\bar F_{X_1}(t),
\end{eqnarray*}
where the second and third inequalities follow from $(i)$ and $(iii)$. Thus, on using the above inequality, we get $k_2(t)\geq 0$. Hence, the result is proved.$\hfill\Box$
\\\hspace*{0.3 in}The following corollary immediately follows from Theorem~\ref{th32}.
\begin{c1}
Let $\{\om_i(\cdot),g_i(\cdot)\}$ be the set of model functions for $X_i$, $i=1,2,\dots,n$, and $\{\ga(\cdot),h(\cdot)\}$ be that for $Y$.  Suppose that $(i)$ and $(ii)$, and $(iii)$ or $(iv)$ hold.
\begin{itemize}
\item [$(i)$] $X_1\leq_{hr} X_2\leq_{hr}\dots\leq_{hr}X_n$ and $g_1(\al)\leq g_2(\al)\leq \dots\leq g_n(\al)$, for all $0\leq \al\leq 1$.
\item [$(ii)$] $Y$ has log-concave survival function, and $u-\ga (u)$ and $\ga(u)-h(1-\al)u$ are increasing in $u\geq 0$, for all $0\leq \al\leq 1$.
\item [$(iii)$] Let $n$ be an even integer. Further, $X_1,X_3,\dots,X_{n-1}$ or $X_2,X_4,\dots,X_{n}$ have log-concave (resp. log-convex) survival functions, and $g_n(\al )u\leq\om_1(u)\leq \om_2(u)\leq\dots\leq \om_n(u)$ (resp. $g_n(\al )u\leq\om_n(u)\leq \om_{n-1}(u)\leq\dots\leq \om_1(u)$), for all $u\geq 0$ and $0\leq \al\leq 1$.
\item [$(iv)$] Let $n$ be an odd integer. Further, $X_1,X_3,\dots,X_{n}$ or $X_2,X_4,\dots,X_{n-1}$ have log-concave (resp. log-convex) survival functions, and $g_n(\al )u\leq\om_1(u)\leq \om_2(u)\leq\dots\leq \om_n(u)$ (resp. $g_n(\al )u\leq\om_n(u)\leq \om_{n-1}(u)\leq\dots\leq \om_1(u)$), for all $u\geq 0$ and $0\leq \al\leq 1$.
\end{itemize}
Then, $U_1\geq_{st}U_2\geq_{st}\dots\geq_{st}U_n$.$\hfill\Box$
\end{c1}
\hspace*{0.3 in}The following theorem shows that a similar result as in Theorem~\ref{th32} holds under some weaker conditions whenever both $X_1$ and $X_2$ have the same set of model functions.
\begin{t1}\label{th34}
Let both $X_1$ and $X_2$ have the same set of model functions given by $\{\om(\cdot),g(\cdot)\}$, and $Y$ have the model function given by $\{\ga(\cdot),h(\cdot)\}$. Assume that $\om(u)\geq g(\al)u$, for all ~$0\leq \al\leq 1$ and $u\geq 0$. If $X_1\leq_{hr}X_2$ then $U_1\geq_{st}U_2$.
\end{t1}
{\bf Proof:} Writing $\zeta(t)=\bar F_{U_1}(t)-\bar F_{U_2}(t)$, we have
\begin{eqnarray}\label{eq1001}
\zeta(t)=k_3(t)+k_4(t),
\end{eqnarray}
where
\begin{eqnarray*}
k_3(t)&=&\bar F_Z(t)\bar F_{Y}(h(1-\al)t)\left[\bar F_{X_2}(t)\bar F_{X_1}(g(\al)t) -\bar F_{X_1}(t)\bar F_{X_2}(g(\al)t)\right]
\\&&+\int\limits_0^t \bar F_Z(t)\bar F_Y(h(1-\al)u)\frac{\bar F_{Y}(t-u+\ga(u))}{\bar F_{Y}(\ga(u))}g(\al)\left[\bar  F_{X_2}(t)f_{X_1}(g(\al)u)\right.
\\&&\left.-\bar F_{X_1}(t)f_{X_2}(g(\al)u)\right]du
\end{eqnarray*}
and
\begin{eqnarray*}
k_4(t)=&&\int\limits_0^t \bar F_Z(t)h(1-\al)f_Y(h(1-\al)u)\left[\frac{\bar F_{X_1}(t-u+\om(u))}{\bar F_{X_1}(\om(u))}\bar F_{X_1}(g(\al)u)\bar F_{X_2}(t)\right.
\\&&\left.-\frac{\bar F_{X_2}(t-u+\om(u))}{\bar F_{X_2}(\om(u))}\bar F_{X_2}(g(\al)u)\bar F_{X_1}(t)\right]du.
\end{eqnarray*}
Since, $X_1\leq_{hr}X_2$, we have, for all $0\leq u\leq t<\infty$,
\begin{eqnarray}\label{eq31}
\bar F_{X_2}(t)\bar F_{X_1}(g(\al)t) -\bar F_{X_1}(t)\bar F_{X_2}(g(\al)t)\geq 0,
\end{eqnarray}
\begin{eqnarray}\label{eq34}
\bar F_{X_2}(t)\bar F_{X_1}(g(\al)u) -\bar F_{X_1}(t)\bar F_{X_2}(g(\al)u)\geq 0,
\end{eqnarray}
and
\begin{eqnarray}\label{eq32}
r_{X_1}(g(\al)u)\geq r_{X_2}(g(\al)u).
\end{eqnarray}
On using (\ref{eq34}) and (\ref{eq32}), we have
\begin{eqnarray}\label{eq33}
\bar  F_{X_2}(t)f_{X_1}(g(\al)u)-\bar F_{X_1}(t)f_{X_2}(g(\al)u)\geq 0.
\end{eqnarray}
Thus, on using (\ref{eq31}) and (\ref{eq33}), we get that $k_3(t)\geq 0$.
Further, $X_1\leq_{hr}X_2$ and $\om(u)\geq g(\al)u$ imply that
\begin{eqnarray}\label{eq35}
\bar F_{X_1}(t-u+\om(u))\bar F_{X_2}(t)-\bar F_{X_2}(t-u+\om(u))\bar F_{X_1}(t)\geq 0
\end{eqnarray}
and
\begin{eqnarray}\label{eq36}
\frac{\bar F_{X_1}(g(\al)u)}{\bar F_{X_1}(\om(u))}\geq \frac{\bar F_{X_2}(g(\al)u)}{\bar F_{X_2}(\om(u))}.
\end{eqnarray}
Thus, on using (\ref{eq35}) and (\ref{eq36}), we have that $k_4(t)\geq 0$. Hence, the result is proved.$\hfill\Box$
\\\hspace*{0.3 in}The following corollary immediately follows from the avobe theorem.
\begin{c1}
Let all $X_1,X_2,\dots,X_n$ have the same set of model functions given by $\{\om(\cdot),g(\cdot)\}$, and $Y$ have the model function given by $\{\ga(\cdot),h(\cdot)\}$. Assume that $\om(u)\geq g(\al)u$, for all ~$0\leq \al\leq 1$ and $u\geq 0$. If $X_1\leq_{hr}X_2\leq_{hr}\dots\leq_{hr}X_n$ then  $U_1\geq_{st}U_2\geq_{st}\dots\geq_{st}U_n$.$\hfill\Box$
\end{c1}
\hspace*{0.3 in}In the following theorem we show that the best strategy to get the optimal parallel system is to allocate the redundancy with the stochastically strongest component of the system (see Theorem 2.4.)
\begin{t1}\label{th33}
Let $\{\om_1(\cdot),g_1(\cdot)\}$, $\{\om_2(u),g_2(\cdot)\}$ and $\{\ga(\cdot),h(\cdot)\}$ be the sets of model functions for $X_1$, $X_2$ and $Y$, respectively. Assume that $X_1\geq_{rhr}X_2$. Suppose that the following conditions hold.
\begin{itemize}
\item [$(i)$] $\om_1(u)=g_1(\al)u\leq g_2(\al)u=\om_2(u)$ for all $u\geq 0$ and $0\leq \al\leq 1$.
\item [$(ii)$] $Y$ has log-concave survival function, and $u-\ga (u)$ and $\ga(u)-h(1-\al)u$ are increasing in $u\geq 0$, for all $0\leq \al\leq 1$.
\end{itemize}
Then, $V_1\geq_{st}V_2$.
\end{t1}
{\bf Proof:} Note that
\allowdisplaybreaks{
\begin{eqnarray*}
\bar F_{X_{1}\oplus Y}(t)&=&\bar F_{X_1}(g_1(\al)t)\bar F_Y(h(1-\alpha)t)
\\&&+\int\limits_0^t \bar F_{X_1}(t-u+\om_1(u))dF_Y(h(1-\al)u)
\\&&+\int\limits_0^t \frac{\bar F_Y(t-u+\ga(u))}{\bar F_Y(\ga(u))}\bar F_Y(h(1-\al)u)dF_{X_1}(g_1(\al)u)
\\&=&\bar F_Y(h(1-\alpha)t)
\\&&+\int\limits_0^t \bar F_{X_1}(t-u+\om_1(u))dF_Y(h(1-\al)u)
\\&&-\int\limits_0^t F_{X_1}(g_1(\al)u)\;d\left[ \frac{\bar F_Y(t-u+\ga(u))}{\bar F_Y(\ga(u))}\bar F_Y(h(1-\al)u)\right],
  \end{eqnarray*}}
  which gives
  \allowdisplaybreaks{
  \begin{eqnarray*}
 F_{X_{1}\oplus Y}(t)&=& F_Y(h(1-\alpha)t)
\\&&-\int\limits_0^t \bar F_{X_1}(t-u+\om_1(u))dF_Y(h(1-\al)u)
\\&&+\int\limits_0^t F_{X_1}(g_1(\al)u)\;d\left[ \frac{\bar F_Y(t-u+\ga(u))}{\bar F_Y(\ga(u))}\bar F_Y(h(1-\al)u)\right]
\\&=&\int\limits_0^t  F_{X_1}(t-u+\om_1(u))dF_Y(h(1-\al)u)
\\&&+\int\limits_0^t F_{X_1}(g_1(\al)u)\;d\left[ \frac{\bar F_Y(t-u+\ga(u))}{\bar F_Y(\ga(u))}\bar F_Y(h(1-\al)u)\right].
  \end{eqnarray*}}
 Writing $\Im_2(t)=F_{V_2}(t)-F_{V_1}(t)$, we have
\begin{eqnarray*}
\Im_2(t)=s_1(t)+s_2(t),
\end{eqnarray*}
where
\begin{eqnarray*}
s_1(t)&=&\int\limits_0^t F_W(t)\left[F_{X_1}(t) F_{X_2}(t-u+\om_2(u))-F_{X_2}(t) F_{X_1}(t-u+\om_1(u)\right]dF_Y(h(1-\al)u)
\end{eqnarray*}
and
\begin{eqnarray*}
s_2(t)=\int\limits_0^t F_W(t)\left[F_{X_1}(t)F_{X_2}(g_2(\al)u)-F_{X_2}(t)F_{X_1}(g_1(\al)u)\right]d\left[ \frac{\bar F_Y(t-u+\ga(u))}{\bar F_Y(\ga(u))}\bar F_Y(h(1-\al)u)\right].
\end{eqnarray*}
Since, $X_1\geq_{rhr}X_2$ and $\om_1(u)=g_1(\al)u\leq g_2(\al)u=\om_2(u)$ we have, for all $u\in[0,t]$,
\begin{eqnarray}\label{eq8}
F_{X_1}(t) F_{X_2}(t-u+\om_2(u))-F_{X_2}(t) F_{X_1}(t-u+\om_1(u)\geq 0
\end{eqnarray}
and
\begin{eqnarray}\label{eq9}
F_{X_1}(t)F_{X_2}(g_2(\al)u)-F_{X_2}(t)F_{X_1}(g_1(\al)u)\geq 0.
\end{eqnarray}
Again, condition $(ii)$ implies that (see proof of Theorem~\ref{th31})
\begin{eqnarray}\label{eq10}
 \frac{\bar F_Y(t-u+\ga(u))}{\bar F_Y(\ga(u))}\bar F_Y(h(1-\al)u\;\;\text{is increasing in }u\geq 0.
\end{eqnarray}
Thus, on using (\ref{eq8}), (\ref{eq9}) and (\ref{eq10}) we get that $s_1(t)\geq 0$ and $s_2(t)\geq 0$, and hence the result follows.$\hfill\Box$
\\\hspace*{0.3 in}As a consequence of the above theorem we have the following corollary.
\begin{c1}
Let $\{\om_i(\cdot),\om(\cdot)\}$ be the set of model functions for $X_i$, $i=1,2,\dots,n$, and $\{\ga(\cdot),h(\cdot)\}$ be that for $Y$. Assume that $X_1\geq_{rhr} X_2\geq_{rhr}\dots\geq_{rhr}X_n$. Suppose that the following conditions hold.
\begin{itemize}
\item [$(i)$] $\om_1(u)=g_1(\al)u\leq g_2(\al)u=\om_2(u)\leq\dots \leq g_n(\al)u=\om_n(u)$, for all $u\geq 0$ and $0\leq \al\leq 1$.
\item [$(ii)$] $Y$ has log-concave survival function, and $u-\ga (u)$ and $\ga(u)-h(1-\al)u$ are increasing in $u\geq 0$, for all $0\leq \al\leq 1$.
\end{itemize}
Then, $V_1\geq_{st}V_2\geq_{st}\dots\geq_{st}V_n$.$\hfill\Box$
\end{c1}
\subsection*{Concluding Remarks}\label{se3}
In this paper, we have considered general load-sharing series and parallel systems. We have shown that, for a load-sharing series (resp. parallel) system, the best strategy to get the optimal system (in the sense of usual stochastic order) is to allocate the redundant component to the weakest (resp. strongest) original component of the system. We have studied the proposed results under cumulative exposer model as well as in a general scenario. The accelerated lifetime model and the virtual age model are used in order to calculate the reliability function of a general load-sharing system. As to the best of our knowledge, there are no results in the literature that deal with allocation strategies for general load-sharing systems, our study might be the first step in this direction. We have considered only series and parallel systems. The study of different allocation strategies for one or more redundant components in a general load-sharing system (for example, $k$-out-of-$n$ system, coherent system) can constitute a topic for the future research.
\subsection*{Acknowledgements:}
\hspace*{0.2 in}Nil Kamal Hazra sincerely acknowledges the financial support from the University of the Free State, South Africa.

\end{document}